\documentclass[aps,prm,superscriptaddress,amsfonts,amsmath,amssymb,showpacs,floatfix,reprint]{revtex4-1}

\usepackage{bm}
\usepackage{graphicx}
\usepackage{amsmath}
\usepackage{amstext}
\usepackage{amssymb}
\usepackage{amsfonts}
\usepackage{amsbsy}
\usepackage{verbatim}
\usepackage[usenames,dvipsnames,svgnames,table]{xcolor}
\usepackage{multirow}
\usepackage{gensymb}
\usepackage{textcomp}
\usepackage{enumitem}
\usepackage{comment}
\usepackage{fancyhdr}
\usepackage[version=3]{mhchem}
\usepackage[colorlinks=true, urlcolor=Blue, linkcolor=Blue, citecolor=Blue, pdftex]{hyperref}
\usepackage{times}
\usepackage{nicefrac}

\newcommand{\be}{\begin{equation}}
\newcommand{\ee}{\end{equation}}

\newcommand{\bee}{\begin{equation*}}
\newcommand{\eee}{\end{equation*}}

\newcommand{\1}{\hspace*{-1pt}}

\begin{document}
\title{Stability of the spiral spin liquid in MnSc$_2$S$_4$}
\author{Yasir Iqbal}
\email[]{yiqbal@physics.iitm.ac.in}
\affiliation{Department of Physics, Indian Institute of Technology Madras, Chennai 600036, India}
\author{Tobias M\"uller}
\affiliation{Institute for Theoretical Physics and Astrophysics, Julius-Maximilian's University of W\"urzburg, Am Hubland, D-97074 W\"urzburg, Germany}
\author{Harald O. Jeschke}
\affiliation{Research Institute for Interdisciplinary Science, Okayama University, 3-1-1 Tsushima-naka, Kita-ku, Okayama 700-8530, Japan}
\author{Ronny Thomale}
\affiliation{Institute for Theoretical Physics and Astrophysics, Julius-Maximilian's University of W\"urzburg, Am Hubland, D-97074 W\"urzburg, Germany}
\author{Johannes Reuther}
\affiliation{Dahlem Center for Complex Quantum Systems and Institut f\"ur Theoretische Physik, Freie Universit\"{a}t Berlin, Arnimallee 14, 14195 Berlin, Germany}
\affiliation{Helmholtz-Zentrum f\"{u}r Materialien und Energie, Hahn-Meitner-Platz 1, 14019 Berlin, Germany} 

\date{\today}

\begin{abstract}
We investigate the stability of the spiral spin liquid phase in MnSc$_2$S$_4$ against thermal and quantum fluctuations as well as against perturbing effects of longer-range interactions. Employing {\it ab initio} density functional theory (DFT) calculations we propose a realistic Hamiltonian for MnSc$_2$S$_4$, featuring second ($J_2$) and third $(J_3)$ neighbor Heisenberg interactions on the diamond lattice that are considerably larger than previously assumed. We argue that the combination of strong $J_2$ and $J_3$ couplings reproduces the correct magnetic Bragg peak position measured experimentally. Calculating the spin-structure factor within the pseudofermion functional-renormalization group technique we find that close to the magnetic phase transition the sizable $J_3$ couplings induce a strong spiral selection effect, in agreement with experiments. With increasing temperature the spiral selection becomes weaker such that in a window around three to five times the ordering temperature an approximate spiral spin liquid is realized in MnSc$_2$S$_4$.
\end{abstract}
\maketitle

\section{Introduction}
If magnetic frustration is sufficiently strong, a spin system may evade spontaneous symmetry breaking at low temperatures and instead form a highly entangled state where the spins fluctuate in a cooperative manner. This so-called spin liquid state generally exists in two different flavors: the quantum~\cite{anderson73,balents10,savary16} and the classical spin liquid~\cite{ramirez99,bramwell01,bergmann07,gao16}. The first case preferably occurs for small quantum spins in combination with frustrated lattice geometries and/or anisotropic interactions where quantum fluctuations may reach the size of the local spin magnitude thus hindering the system from developing magnetic long-range order.
In the second case, spin liquid-like behavior even survives in the complete absence of quantum fluctuations such as for classical ($S\to\infty$) spins. The suppression of long-range magnetic order now relies on a macroscopic degeneracy of classical ground states through which the system fluctuates collectively, thus justifying the notion of a classical spin liquid. Paradigmatic examples are pyrochlore spin-ice systems~\cite{ramirez99,bramwell01}, where at zero temperature an ice rule (e.g., the famous two-in-two-out rule) imposes local constraints on possible spin states. Since these rules leave the ground-state spin configuration underdetermined, the system maintains a macroscopic (extensive) classical degeneracy~\cite{pauling35}.

\begin{figure}[t]
\includegraphics[width=0.99\linewidth]{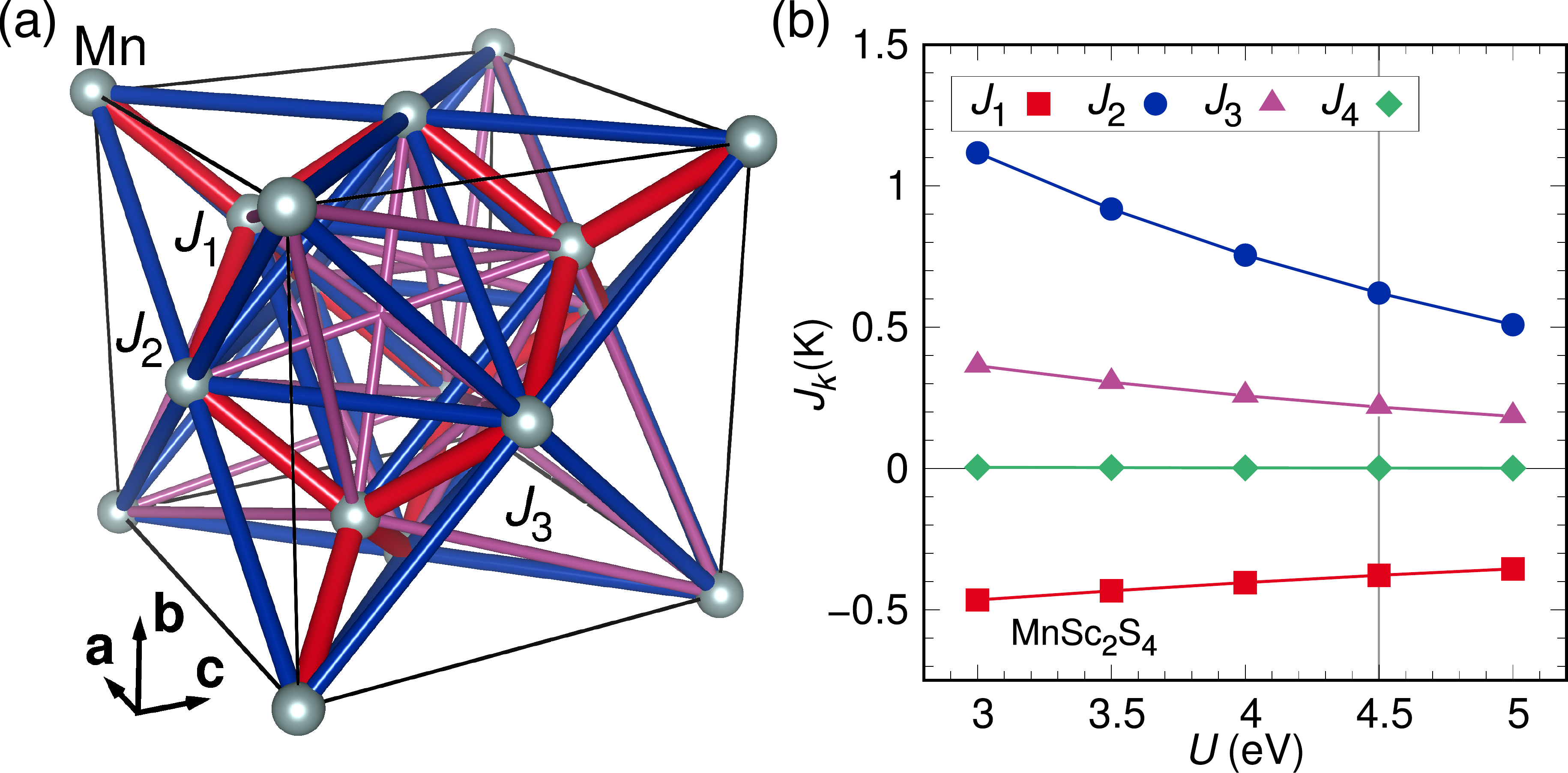}
\caption{(a) Cubic unit cell of the diamond lattice with first ($J_1$), second ($J_2$), and third ($J_3$) neighbor couplings. (b) Couplings $J_{1}$\textendash$J_{4}$ from DFT as a function of the Hubbard $U$ interaction. The vertical line indicates the exchange couplings investigated in the main text. \label{fig1}}
\end{figure}

Interestingly, for certain lattice geometries and special arrangements of frustrating interactions, classical spin liquids even exist without a local ice-rule constraint. This rare situation is realized on the three-dimensional diamond lattice [Fig.~\ref{fig1}(a)] with first ($J_1$) and second ($J_2$) neighbor Heisenberg interactions when $\frac{J_2}{|J_1|}>\frac{1}{8}$ and $J_2$ is antiferromagnetic~\cite{bergmann07,lee08,savary11,buessen18}. The competing interactions force the system into classical coplanar spin-spirals. Remarkably, the ground state is formed from a highly degenerate set of such spirals where the corresponding wave vectors $\mathbf{q}$ occupy a closed surface in reciprocal space (note that a similar scenario also occurs on the two-dimensional honeycomb lattice~\cite{fouet01,mulder10,baez17}). Due to the cooperative motion of spins through the degenerate manifold of spirals, this state has been dubbed a {\it spiral spin liquid}.

Spiral spin liquids are generally very fragile to perturbations of various different types. Any finite additional term in the Hamiltonian such as third neighbor couplings $J_3$ or dipolar interactions typically selects specific spirals out of the degenerate manifold and consequently generates long-range magnetic order. Even in the absence of such perturbations, a lifting of the degeneracy takes place due to thermal fluctuations, i.e., a finite temperature transition into a magnetically ordered state is induced by an entropic ``order-by-disorder'' selection~\cite{villain80} of spirals. As has been found in Ref.~\cite{bergmann07}, by varying $\frac{J_2}{|J_1|}>\frac{1}{8}$ the system goes through a sequence of different magnetic phases. While strictly speaking this effect destroys spiral spin liquids at any finite temperature, an approximate version of this state may still survive in a temperature range {\it above} the transition where the thermal selection is not yet active. Finally, quantum fluctuations at large but finite spin magnitudes have been found to induce an order-by-disorder effect similar to thermal fluctuations~\cite{buessen18}.

Currently, the most promising material to approximately realize a spiral spin liquid is the A-site spinel MnSc$_2$S$_4$~\cite{fritsch04,giri05,krimmel06,kalvius06,muecksch07,gao16} where spin\textendash5/2 Mn$^{2+}$ ions occupy the sites of a diamond lattice. At $\sim$$2.9$ K which is well below the Curie-Weiss temperature of $|\Theta_{\text{CW}}|=23$ K~\cite{fritsch04} but still inside the paramagnetic phase of this compound (which survives down to $\sim$$2.3$ K~\cite{fritsch04,krimmel06,kalvius06,muecksch07,gao16}) neutron scattering directly observes surface-like scattering profiles in momentum space, reminiscent of a spiral spin liquid~\cite{gao16}. From the radius of this surface a coupling ratio of $\frac{J_2}{|J_1|}=0.85$ has been determined~\cite{bergmann07} (where $J_1$ is ferromagnetic). The measured spin-structure factor is not evenly distributed on the spiral surface but shows higher intensities for spirals with wave vectors $\mathbf{q}\sim2\pi(0.75,0.75,0)$ and symmetry-related positions~\cite{krimmel06,muecksch07,gao16}. This spiral selection turns into real magnetic long-range order below $T_{\rm c}=2.3$ K~\cite{krimmel06,muecksch07} (other works report slightly smaller values of $T_{\rm c}\approx2.1$ K~\cite{fritsch04,kalvius06,gao16}). It is worth emphasizing that this peak position does not coincide with the thermal selection predicted in Ref.~\cite{bergmann07} but rather points towards the presence of longer-range $J_3$ interactions.

\begin{figure*}
\includegraphics[width=0.99\linewidth]{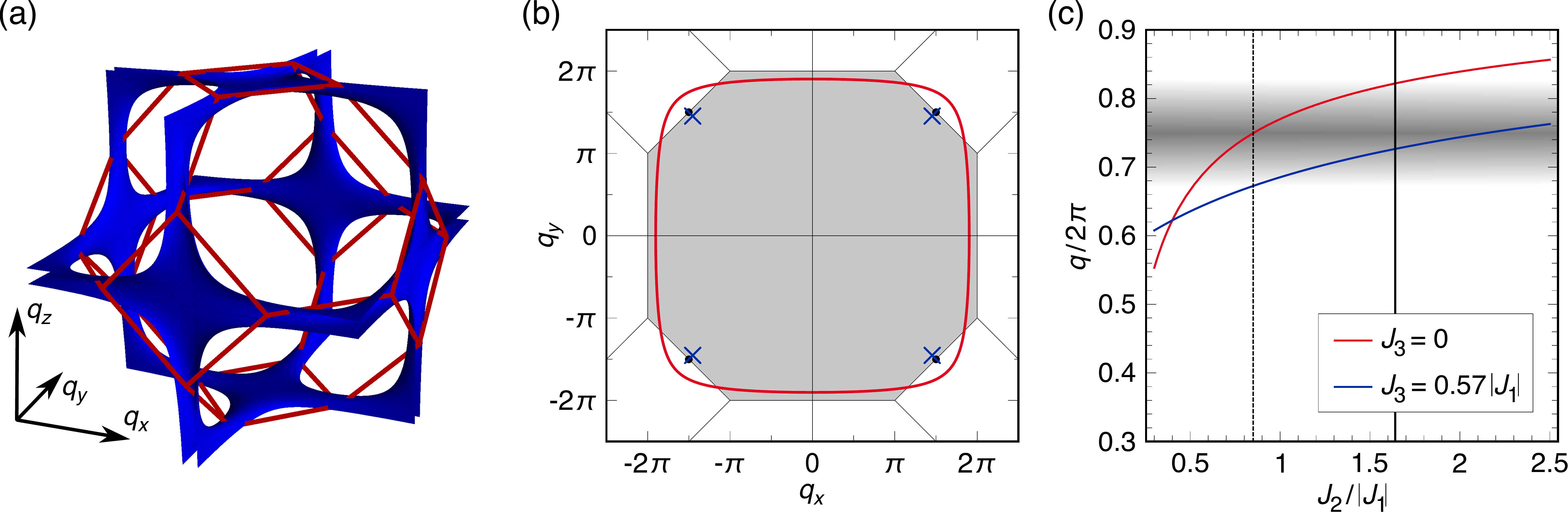}
\caption{(a) Classical spin-spiral surface in the extended Brillouin zone formed by the wave-vectors of the degenerate spiral ground states of the model with $J_2/|J_1|=1.64$ and $J_3=0$, (b) The classical spin-spiral surface in the $q_x$\textendash$q_y$ plane for $J_2/|J_1|=1.64$ and $J_3=0$ (red line). The blue crosses indicate the Bragg peak position for an additional third neighbor coupling $J_3/|J_1|=0.57$. Black dots highlight the measured magnetic order at $\mathbf{q}\sim2\pi(0.75,0.75,0)$. (c) Red: Size of the spiral surface [given by the intersection with the line $(q,q,0)$] as a function of $J_2$ and for $J_3=0$. Blue: $(q,q,0)$ position of the ordering wave vector for $J_3/|J_1|=0.57$. Vertical full (dashed) lines indicate the coupling ratios $J_2/|J_1|=1.64$ ($J_2/|J_1|=0.85$~\cite{bergmann07}). The shaded area marks the position and width of the measured magnetic Bragg peak $\mathbf{q}\sim2\pi(0.75,0.75,0)$~\cite{gao16}. 
\label{fig2}}
\end{figure*}

This article complements recent experimental works by theoretically investigating the fate of the spiral spin liquid when assuming a realistic model for MnSc$_2$S$_4$. To this end, we first employ {\it ab initio} density functional theory (DFT) calculations to determine the microscopic Hamiltonian of this compound. We then treat the resulting model within the pseudofermion functional-renormalization group (PFFRG) method~\cite{reuther10} which is capable of resolving the effects of thermal and quantum fluctuations, and we clarify the role of third neighbor $J_3$ interactions. In particular, we investigate to which degree the spiral spin liquid phase in MnSc$_2$S$_4$ remains stable under such perturbations and compare the $\mathbf{q}$\textendash space resolved magnetic susceptibility with neutron scattering experiments. Our main results are summarized as follows: (i) We find that the $J_2$ and $J_3$ interactions are both considerably larger than previously assumed~\cite{bergmann07}. (ii) Close to the magnetic phase transition but still inside the paramagnetic regime the spin correlations are dominated by $J_3$ couplings which induce a pronounced selection of spirals with wave vectors $\mathbf{q}\approx2\pi(0.72,0.72,0)$, in excellent agreement with experiments. (iii) We identify a temperature regime around $3T_{\rm c}$ to $5T_{\rm c}$ where the spiral selection due to $J_3$ couplings is suppressed such that the system realizes an approximate spiral spin liquid. (iv) PFFRG calculations for our model Hamiltonian reproduce the measured spin structure factor for MnSc$_2$S$_4$ with remarkable accuracy.

The paper is structured as follows: In Sec.~\ref{sec:methods}, we describe the DFT and PFFRG methods, and provide details of the calculations. In Sec.~\ref{sec:mm} we discuss the model Hamiltonian determined from DFT and compare our exchange couplings with those of the previously proposed model. We also discuss the physical implications of these new couplings for the corresponding classical model employing the Luttinger-Tisza method. Section~\ref{sec:PFFRG} contains the results obtained from the PFFRG calculations for the newly proposed Hamiltonian, which are also compared and contrasted with those obtained for the previously proposed model. Finally, in Sec.~\ref{sec:discussions} we summarize and discuss our findings, and give concluding remarks.    

\begin{table}[t]
\setlength{\tabcolsep}{2.27pt}
\centering
\begin{tabular}{lllllc}
 \hline \hline
       \multicolumn{1}{c}{$U$ (eV)}
    & \multicolumn{1}{c}{$J_{1}$ (K)}
    & \multicolumn{1}{c}{$J_{2}$ (K)}
    & \multicolumn{1}{c}{$J_{3}$ (K)}
    & \multicolumn{1}{c}{$J_{4}$ (K)}
    & \multicolumn{1}{c}{$\Theta_{\rm CW}$ (K)}  \\ \hline
       
\multirow{1}{*}{3.0} & $-0.465(2)$ & 1.117(1) & 0.364(1) & 0.0039(6) & $-46$ \\ 
\multirow{1}{*}{3.5} & $-0.433(2)$ & 0.918(1) & 0.305(1) & 0.0029(5) & $-38$ \\
\multirow{1}{*}{4.0} & $-0.404(1)$ & 0.755(1) & 0.257(1) & 0.0022(4) & $-31$ \\
\multirow{1}{*}{$\mathbf{4.5}$} & $\mathbf{-0.378(1)}$ & $\mathbf{0.621(1)}$ & $\mathbf{0.217(1)}$ & $\mathbf{0.0015(3)}$ & $\mathbf{-25}$ \\
\multirow{1}{*}{5.0} & $-0.356(1)$ & 0.509(1) & 0.184(1) & 0.0009(3) & $-20$ \\ \hline 
\end{tabular}
\caption{Exchange couplings of MnSc$_2$S$_4$ calculated within GGA+$U$ at $J_{\rm H}=0.76$~eV and $6\times 6\times 6$ $q$-points. The parameters corresponding to $U = 4.5$ eV (marked in bold) are used for the PFFRG simulations [see also Fig.~\ref{fig1}(b)].}
\label{tab:DFT}
\end{table}

\section{Methods}\label{sec:methods}
We base our calculations on the cubic spinel structure determined by
neutron powder diffraction at $T=1.6$~K~~\cite{krimmel06}. The
Mn$^{2+}$ ions form a diamond lattice as shown in
Fig.~\ref{fig1}(a). We use an energy mapping technique to determine
the most important exchange interactions in
MnSc$_2$S$_4$~\cite{Jeschke2013,Guterding2016,Iqbal2017}. For
this purpose we construct a $2\times 2\times 1$ supercell of the
original primitive cell containing two Mn$^{2+}$ ions; in $P\,m$ space
group, this supercell has eight inequivalent Mn sites allowing for 20
distinct spin configurations. This allows us to determine the first
four exchange couplings, extending up to a Mn\textendash Mn distance of
10.6~{\AA}. We perform density functional theory calculations with the
all electron full potential local orbital (FPLO)~\cite{Koepernik1994}
basis set and generalized gradient approximation
(GGA)~\cite{Perdew1996} exchange correlation functional, accounting
for the strong correlations on the Mn $3d$ orbitals by a
GGA+$U$~\cite{Liechtenstein95} correction. The Hunds rule coupling for
Mn $3d$ was fixed at $J_{\rm H}=0.76$~eV~~\cite{Mizokawa1996}. The result of
fitting the DFT total energies against the Heisenberg Hamiltonian
\begin{equation}
\mathcal{\hat{H}}=\sum_{k=1}^4\sum_{\langle ij\rangle_k}J_k\mathbf{\hat{S}}_i\cdot\mathbf{\hat{S}}_j\,,\label{ham}
\end{equation}
where $\langle ij\rangle_k$ denotes pairs of $k$th neighbor sites on the diamond lattice, is shown in Fig.~\ref{fig1}(b) and Table~\ref{tab:DFT} for five values of the interaction strength $U$. Note that each pair of sites in the summation of Eq.~(\ref{ham}) is accounted for only once, i.e., we adopt the convention of single counting of bonds. As explained below, the value of $U$ is fixed by the experimentally observed Curie-Weiss temperature $\Theta_{\text{CW}}$.

The spin Hamiltonian from DFT is taken as an input for the PFFRG method~\cite{reuther10}. To treat this model within standard many-body techniques, the PFFRG first expresses the spin operators in terms of Abrikosov pseudofermions~\cite{abrikosov65}. The implementation of the local spin\textendash5/2 moments is performed as described in Ref.~\cite{baez17} where multiple copies of spin\textendash1/2 degrees of freedom effectively realize spins with larger magnitudes. The resulting fermionic Hamiltonian is then investigated using the well-developed FRG method~\cite{metzner12,platt13}, which calculates the evolution of $m$-particle vertices as a function of an RG parameter $\Lambda$. Effectively, the vertex flow takes into account leading diagrammatic contributions in $1/S$~\cite{baez17} and $1/N$~\cite{buessen17,roscher17,Rueck-2018}, such that classical spin correlations and quantum fluctuations (described in large $S$ and large $N$ approaches, respectively) are both faithfully captured. After its initial development in two dimensions~\cite{reuther10}, the PFFRG was further refined and applied to various models of frustrated magnetism including multilayer, and, eventually, three-dimensional magnets~\cite{Balz16,iqbal16,iqbal15,reuther11,reuther11_2,reuther11_3,singh12,reuther14,suttner14,iqbal16_2,iqbal16_3,buessen16,Iqbal2017,buessen17,roscher17,baez17,buessen18,keles18,Chillal-2017,Iqbal-2018,Hering-2018}. The finite-size approximation in the PFFRG amounts to limiting the real-space distance of spin correlations, which in our calculations extends over $12$ nearest-neighbor lattice spacings, corresponding to a correlation volume of $1963$ sites. Likewise, the continuous frequency arguments of the vertex functions are approximated by a discrete set of $64$ frequencies. The central physical quantity studied within the PFFRG is the static (zero-frequency) momentum-resolved susceptibility (or spin structure factor) which can be directly compared with experimental neutron scattering data.  

\section{Model Hamiltonian and classical considerations}\label{sec:mm}
We first discuss the exchange couplings $J_k$ in Eq.~(\ref{ham}) determined
from DFT. As shown in Fig.~\ref{fig1}(b), DFT calculates these couplings as a function of the Hubbard onsite interaction $U$. Upon increasing $U$, all couplings decrease but their ratios remain relatively constant. The actual size of $U$ is determined via the known Curie-Weiss temperature $\Theta_{\text{CW}}=-\frac{S(S+1)}{3k_\text{B}}\sum_{k=1}^4 z_k J_k=-23$ K~\cite{fritsch04} (where $z_k$ is the coordination number of the $k$th neighbor bonds). This condition is best fulfilled for $U\approx4.5$ eV, yielding three significant couplings $J_1=-0.378$ K, $J_2=0.621$ K, $J_3=0.217$ K, and $J_4=0.0015$ K. Since $J_4$ is more than an order of magnitude smaller than all other couplings it will be neglected in the ensuing analysis. The small absolute values of the exchange couplings can be understood from the fact that in the diamond lattice of MnSc$_{2}$S$_{4}$ even the nearest-neighbor exchange couplings $J_{1}$ are mediated via rather long Mn-S-Sc-S-Mn superexchange paths. While the exchange couplings of $<1$~K are small, importantly the energy differences that need to be resolved within DFT are not: due to the spin\textendash$5/2$ moments, the energies for the different spin configurations vary in a window of $20$~meV, which is an energy scale that can be comfortably resolved by our highly converged all electron full potential DFT calculations.

\begin{figure*}
\includegraphics[width=1.0\linewidth]{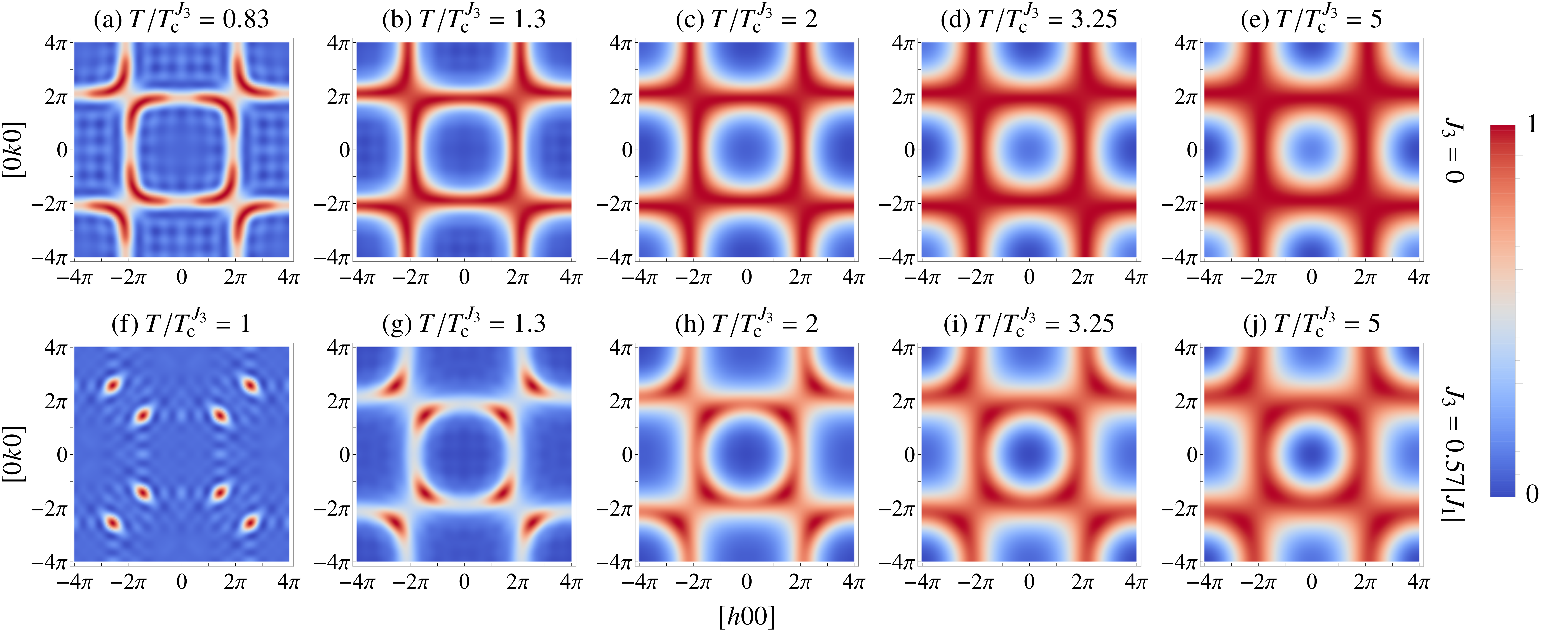}
\caption{The evolution of the spin susceptibility profile in the $q_{x}$\textendash$q_{y}$ plane with temperature for the Heisenberg Hamiltonian of MnSc$_2$S$_4$ as determined from DFT. (a)\textendash(e) Evaluated for a $J_{1}$\textendash$J_{2}$ {\it only} model with $J_{2}/|J_{1}|=1.64$ and $J_{3}=0$; (f)\textendash(j) evaluated for the full $J_{1}$\textendash$J_{2}$\textendash$J_{3}$ Hamiltonian with $J_{2}/|J_{1}|=1.64$ and $J_{3}/|J_{1}|=0.57$. The temperatures are expressed in units of the critical (ordering) temperature $T_{\rm c}^{J_{3}}$ of the full model Hamiltonian with $J_{3}/|J_{1}|=0.57$. Note that in both models, at each temperature, we have rescaled the susceptibility so as to make the minimum and maximum plotted values lie between 0 and 1, which makes prominent the important features characterizing the spiral spin liquid. The absolute values of the maxima can be read off from the temperature evolution of the susceptibility shown in Fig.~\ref{fig10} [see black curve for system size L $=12$]. For each of the above profiles, the variation of the susceptibility along the radial $(q,q,0)$ direction is shown in Fig.~\ref{fig7}.} \label{fig3}
\end{figure*}

The DFT couplings might first appear unexpected because the ratios $\frac{J_2}{|J_1|}=1.64$ and $\frac{J_3}{|J_1|}=0.57$ are considerably larger compared to the values $\frac{J_2}{|J_1|}\approx0.85$ and $\frac{J_3}{|J_1|}\lesssim0.1$ proposed earlier (see Refs.~\cite{bergmann07,lee08}, respectively). These values were obtained from matching calculated and measured inelastic neutron scattering spectra under the assumption that $J_{3}$ is negligible~\cite{gao16}. However, in materials featuring a number of competing interactions, fitting methods are known to be ambiguous (see e.g., Refs.~\cite{Jeschke-2011,Janson-2016}), and thus, DFT based methods provide an  important complementary path towards extraction of couplings allowing for an identification of the relevant Hamiltonian. Indeed, our DFT results reproduce the sign of the nearest- and next-nearest-neighbor exchange couplings of MnSc$_2$S$_4$ proposed earlier~\cite{bergmann07}, furthermore, they refine the previous picture by highlighting the presence of significant $J_{3}$ couplings, which considerably alters our understanding of the mechanism leading to the stabilization of a spiral spin liquid. 

To shed further light on the physical implications of these new couplings, we first treat Eq.~(\ref{ham}) in the classical limit, employing the Luttinger-Tisza method~\cite{luttinger46,luttinger51}. This method aims at calculating the ground state of the corresponding classical Heisenberg Hamiltonian by minimizing the energy given by Eq.~(\ref{ham}), and does so by relaxing the spins' length constraint at each site, however, on the diamond-lattice geometry this soft-spin approach even becomes exact (see Appendix~\ref{sec:LT}). Ignoring $J_3$ for a moment, the $J_1$\textendash$J_2$ only model with $\frac{J_2}{|J_1|}=1.64$ exhibits a spiral surface in momentum space [see Fig.~\ref{fig2}(a)], which cuts through the first Brillouin-zone boundary [see Fig.~\ref{fig2}(b)]. This surface is slightly larger than the one for $\frac{J_2}{|J_1|}=0.85$, where the latter ratio has been determined in Ref.~\cite{bergmann07} to match the measured magnetic Bragg peak position $\mathbf{q}\approx2\pi(0.75,0.75,0)$ for $J_3=0$. Although the spiral surface only undergoes a moderate increase between $\frac{J_2}{|J_1|}=0.85$ and $\frac{J_2}{|J_1|}=1.64$, the DFT couplings first seem to overestimate the ordering wave vector even when the finite Bragg-peak width is taken into account [see Fig.~\ref{fig2}(c)]. The situation changes when $J_3$ couplings are considered. Already an infinitesimally small $J_3$ lifts the degeneracy and selects spirals with $\mathbf{q}=(q,q,0)$ along the surface. For larger (antiferromagnetic) $J_3$ this Bragg-peak position moves inwards in $\mathbf{q}$ space. As shown in Figs.~\ref{fig2}(b) and \ref{fig2}(c), the third neighbor coupling $\frac{J_3}{|J_1|}=0.57$ from DFT indeed shifts the Bragg peak back to $\mathbf{q}=2\pi(0.73,0.73,0)$, in very good agreement with the measured position. As discussed in Ref.~\cite{lee08}, small remaining discrepancies might disappear when incommensurate/commensurate ``lock-in'' transitions are considered.

\section{PFFRG results}\label{sec:PFFRG}
Having argued that our model parameters are generally compatible with the experimental findings, we next investigate to what extent the strong $J_3$ coupling together with thermal and quantum fluctuations destabilize the spiral spin liquid. To this end, we first calculate the spin-structure factor via PFFRG for $\frac{J_2}{|J_1|}=1.64$ and $J_3=0$, where only the effects of thermal and quantum fluctuations lift the spiral degeneracy [see Figs.~\ref{fig3}(a)\textendash\ref{fig3}(e)], and then compare with $\frac{J_3}{|J_1|}=0.57$, to study the influence of additional third neighbor couplings [see Figs.~\ref{fig3}(f)\textendash\ref{fig3}(j), and Figs.~\ref{fig8} and~\ref{fig9}]. In both cases, the spin-structure factor is investigated as a function of the RG parameter $\Lambda$ which has been argued to mimic finite temperatures $T$~\cite{reuther11,iqbal16,buessen16}. Indeed, the conversion factor between the RG scale $\Lambda$ and temperature $T$ evaluates to $\frac{T}{J}=\Big{(}\frac{2\pi S(S+1)}{3}\Big{)}\frac{\Lambda}{J}$. This is determined by comparing the limit of PFFRG where only the RPA diagrams contribute, i.e., a mean-field description, and the conventional spin mean-field theory formulated in terms of temperature $T$ instead of $\Lambda$~\cite{iqbal16,baez17}.

\begin{figure}[t]
\includegraphics[width=1\linewidth]{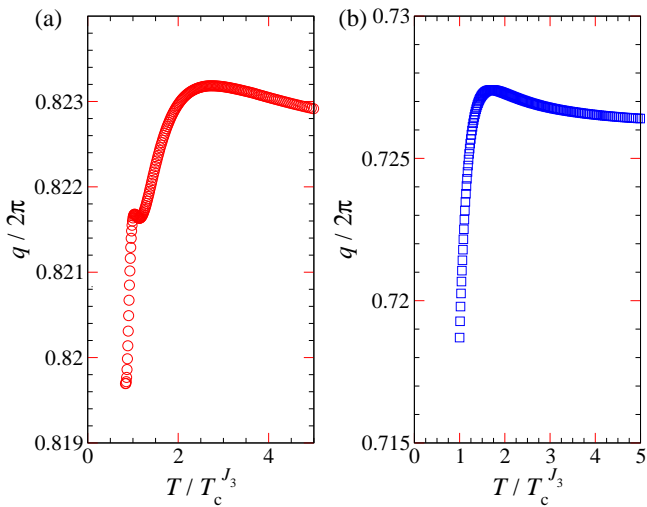}
\caption{The variation with temperature of the peak-position $q$ of the susceptibility maximum along the radial $(q,q,0)$ direction for the model Hamiltonian with $\frac{J_{2}}{|J_{1}|}=1.64$ and (a) $J_{3}=0$ and (b) $J_{3}=0.57|J_{1}|$.}\label{fig4}
\end{figure}
\begin{figure}[b]
\includegraphics[width=1.0\linewidth]{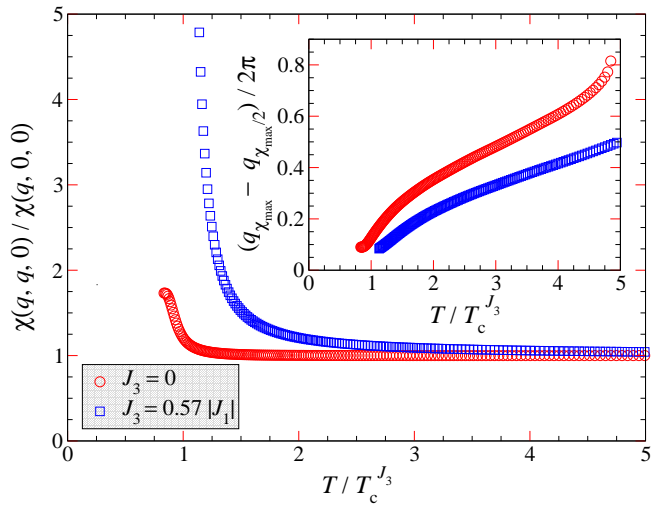}
\caption{The ratio of the susceptibility maxima along the $(q,q,0)$ and $(q,0,0)$ directions shown as a function of temperature $T$. Note that this ratio does not need to diverge when approaching criticality. (Inset) Temperature evolution of the width of the spiral surface along the $(q,q,0)$ direction. The width is defined as the difference of the $q$ values for the maxima and half-maxima of the susceptibility, respectively.}\label{fig5}
\end{figure}

For $J_3=0$ and at the critical RG scale (which corresponds to $\Lambda_{\rm c}^{0}=0.83(1)|J_1|$), the PFFRG detects a sharp spiral contour of strong intensities. Along finite segments centered around $(q,q,0)$ we find somewhat larger (and nearly constant) responses; however, this modulation quickly disappears with increasing temperature (i.e., RG scale $\Lambda$) such that an almost perfect spiral surface appears. Interestingly, the size and shape of the spiral surface remains nearly constant as a function of temperature [see Fig.~\ref{fig4}(a)] while its width increases considerably (see inset of Fig.~\ref{fig5}). Note that due to the missing $J_3$ coupling in Fig.~\ref{fig4}(a) the calculated maximum position $q$ is considerably larger than the experimentally measured wave vector $2\pi(0.75,0.75,0)$ [see also Figs.~\ref{fig2}(b) and \ref{fig2}(c)]; however, the inclusion of a third neighbor coupling $J_3=0.57|J_{1}|$ shifts the peaks to a position very close to the measured value, as discussed below. A more quantitative measure for the intactness of the spiral surface is shown in Fig.~\ref{fig5}, where the ratio of the intensity maximum along the $(q,q,0)$ and along the $(q,0,0)$ direction is plotted. For $J_{3}=0$, this quantity approaches unity, i.e., $\chi(q,q,0)/\chi(q,0,0)\approx1$ at around $\Lambda\approx2 \Lambda_{\rm c}^0$, indicating that the spiral surface quickly recovers. We also note that, compared to our classical Luttinger-Tisza analysis, the location of the spiral surface does not undergo any noticeable changes when including quantum fluctuations.

Switching on the third neighbor coupling $\frac{J_3}{|J_1|}=0.57$ induces a much stronger spiral-selection effect. At criticality, we observe pronounced peaks at $\mathbf{q}=2\pi(0.719,0.719,0)$ [see Fig.~\ref{fig3}(f)], which are found to be shifted slightly inwards compared to the classical wave-vector position $\mathbf{q}=2\pi(0.727,0.727,0)$. The critical RG scale $\Lambda_{\rm c}^{J_{3}}=0.99(1)|J_1|$ is slightly larger compared to the one for $J_3=0$, indicating that third neighbor interactions {\it reduce} the frustration (see Figs.~\ref{fig8} and~\ref{fig9} for a general trend with $J_{3}$). With increasing temperature, the response again becomes more evenly distributed along the spiral surface (see Fig.~\ref{fig5}); however, this intensity smearing occurs more slowly than for $J_3=0$ (see Fig.~\ref{fig8} for susceptibility plots corresponding to different values of $J_{3}$ for fixed $J_{2}/|J_{1}|=1.64$, and Fig.~\ref{fig9} for results with different $J_{3}$ with fixed $J_{2}/|J_{1}|=0.85$~\cite{bergmann07}). 

We now highlight a number of features of our susceptibility data which enable us to establish the existence, stability, and extent of the spiral spin liquid. First, and foremost, a spiral spin liquid is expected to display a near uniform distribution of the susceptibility along a ring-like pattern. To this end, we plot in Fig.~\ref{fig5} the ratio of the susceptibility maxima along $(q,q,0)$ and $(q,0,0)$ directions as a function of temperature. We see that while the ratio starts with a large (diverging) value at $\Lambda_{\rm c}^{J_{3}}$, it slowly converges towards 1. Indeed, at around $\Lambda\approx3\Lambda_{\rm c}^{J_{3}}$ we observe the beginning of a temperature regime where the surface appears relatively intact (note that this temperature reflects a smooth crossover and not a sharp transition). Second, the width of the spiral surface is also seen to decrease upon inclusion of a $J_{3}$ coupling (see inset of Fig.~\ref{fig5}), implying that the response is concentrated within a narrower stripe around the spiral surface compared to the case with $J_{3}=0$, leading to a well-defined and ``intact'' spiral spin liquid. Third, we can obtain a rough estimate for the upper crossover temperature into the spiral spin liquid regime, defined as the temperature where the width of the peaks in the $(q,q,0)$ direction (as shown in the inset of Fig.~\ref{fig5}) equals their separation [where the separation refers to the two peaks which are approximately located at $2\pi(0.75,0.75,0)$ and $2\pi(1.25,1.25,0)$]. Below this temperature, individual spiral surfaces are clearly discernible, which is an important requirement for the realization of a stable spiral spin liquid. For $J_3=0.57|J_1|$ this crossover temperature is roughly given by $T_{\rm crossover} \approx 5 T_\text{c}^{J_3}$, while for $J_3=0$ we find $T_{\rm crossover} \approx 3 T_\text{c}^{J_3}$. These results, taken together, lead to the following approximate phase diagram: (i) Starting from the low-temperature regime, we have for $T/T_{\rm c}^{J_{3}}\leqslant 1$ long-range spiral magnetic order. (ii) For $1<T/T_{\rm c}^{J_{3}}\lesssim3$, we see fingerprints of a ``molten'' spiral order wherein the spectral weight remains concentrated around the ordering wave vectors of the parent spiral order, but the phase is not magnetically long-range ordered. (iii) In the interval $3\lesssim T/T_{\rm c}^{J_{3}}\lesssim5$ we find that not only is the spectral weight nearly uniformly distributed along a spiral surface but also the individual classical spiral spin surfaces are clearly discernible, and the system thus approximately realizes a stable spiral spin liquid. It is worth noting that the temperatures in this window are still much smaller compared to the absolute value of the Curie-Weiss temperature $|\Theta_{\rm CW}|=23$ K. (iv) At higher temperatures $T/T_{\rm c}^{J_{3}}\gtrsim5$, the different spiral surfaces start merging, being no longer individually distinguishable, and the spiral spin liquid becomes unstable towards a high-temperature paramagnet. Most importantly, our PFFRG results indicate that, in a temperature interval of around three to five times
the ordering temperature of $T_{\rm c}=2.3$ K, MnSc$_2$S$_4$ indeed realizes an approximate spiral spin liquid. 

\begin{figure}[t]
\includegraphics[width=1.0\linewidth]{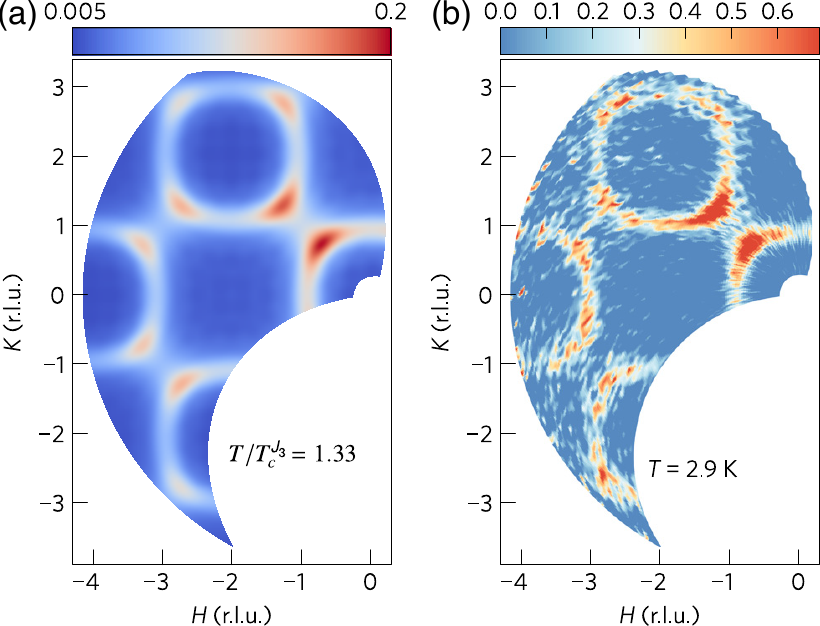}
\caption{Calculated (a) and measured (b) spin-structure factors in the $q_x$\textendash$q_y$ plane for $T/T_{\rm c}^{J_{3}}=1.33$ and $T=2.9$ K, respectively [(b) has been reproduced from Ref.~\cite{gao16}]. The calculated susceptibilities from PFFRG are given in units of $1/|J_{1}|$, while the experimental data are shown in arbitrary units.} \label{fig6}
\end{figure}
\begin{figure}[b]
\includegraphics[width=1\linewidth]{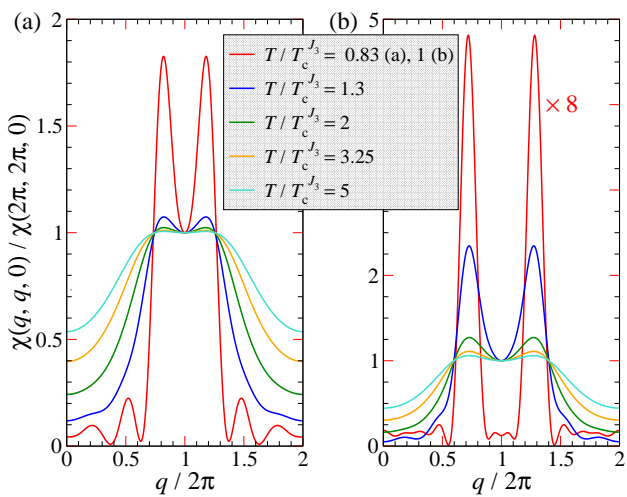}
\caption{The variation of the susceptibility along the radial $(q,q,0)$ direction at different temperatures for a model Hamiltonian with $J_{2}/|J_{1}|=1.64$ and (a) $J_{3}=0$ and (b) $J_{3}=0.57|J_{1}|$.}\label{fig7}
\end{figure}

Finally, to directly assess the quality of our simulations, we compare the measured spin structure-factor at $T=2.9$ K $=1.33T_{\rm c}$~\cite{gao16} with the PFFRG result for the full DFT model at the same RG-scale ratio $\Lambda=1.33\Lambda_{\rm c}^{J_{3}}$. For a proper comparison between theory and experiment, one has to take into consideration the extended orbital structure of the Mn$^{2+}$ magnetic moments as probed by neutron scattering wherein the measured spin structure factor is modulated by a $|\mathbf{q}|$-dependent function \textemdash the so called magnetic form factor~\cite{brown} \textemdash which describes the scattering from single moments (note that the susceptibility profiles in Fig.~\ref{fig3} assume point-like magnetic moments). The magnetic form factor is given by a sum of Gaussian curves with coefficients that can be found in Ref.~\cite{brown}. We have therefore multiplied our PFFRG result with the magnetic form factor of Mn$^{2+}$ ions which leads to a slight decrease of the intensity with increasing $|\mathbf{q}|$. The corresponding susceptibility profile is presented in Fig.~\ref{fig6}. As can be seen, the measured intensity modulation and, in particular, the spiral selection (which is still pronounced at these temperatures) is nicely reproduced by our calculations.

\section{Discussions and conclusions}\label{sec:discussions}
By combining {\it ab initio} DFT and PFFRG calculations we have shown that close to criticality the magnetic ordering process of MnSc$_2$S$_4$ is dominated by a pronounced $(q,q,0)$ spiral selection due to strong $J_3$ couplings, i.e., $J_{3}/|J_{1}|=0.57$, which are significantly larger than previously assumed~\cite{bergmann07}. Yet, as temperature increases, thermal fluctuations largely restore the spiral surface such that an approximate version of a spiral spin liquid is realized at around three to five times the ordering temperature. Interestingly, we find that the $J_3$ coupling is not entirely detrimental to a spiral spin liquid, since the selection induced by such interactions is accompanied by a reduction of the spiral surface's width.

While the Heisenberg couplings considered here determine the momentum structure of the spin correlations, they leave the plane of spiral rotation undetermined. This remaining degeneracy may be further lifted by anisotropic interactions such as dipolar couplings~\cite{lee08,gao16}. However, with a magnitude of a few percent of $J_1$ (Ref.~\cite{gao16} gives an estimate of $\sim0.026$ K on nearest-neighbor bonds) we expect dipolar interactions to become relevant only very close to the ordering transition. On the other hand, {\it below} criticality such couplings might be crucial for explaining the measured multistep ordering process involving sinusoidal collinear, incommensurate, and helical spin orders~\cite{gao16}. Since the PFFRG in its current formulation does not explicitly take into account spontaneous symmetry breaking, an analysis of such phases goes beyond the scope of the present work. 

\begin{figure*}
\includegraphics[width=0.95\linewidth]{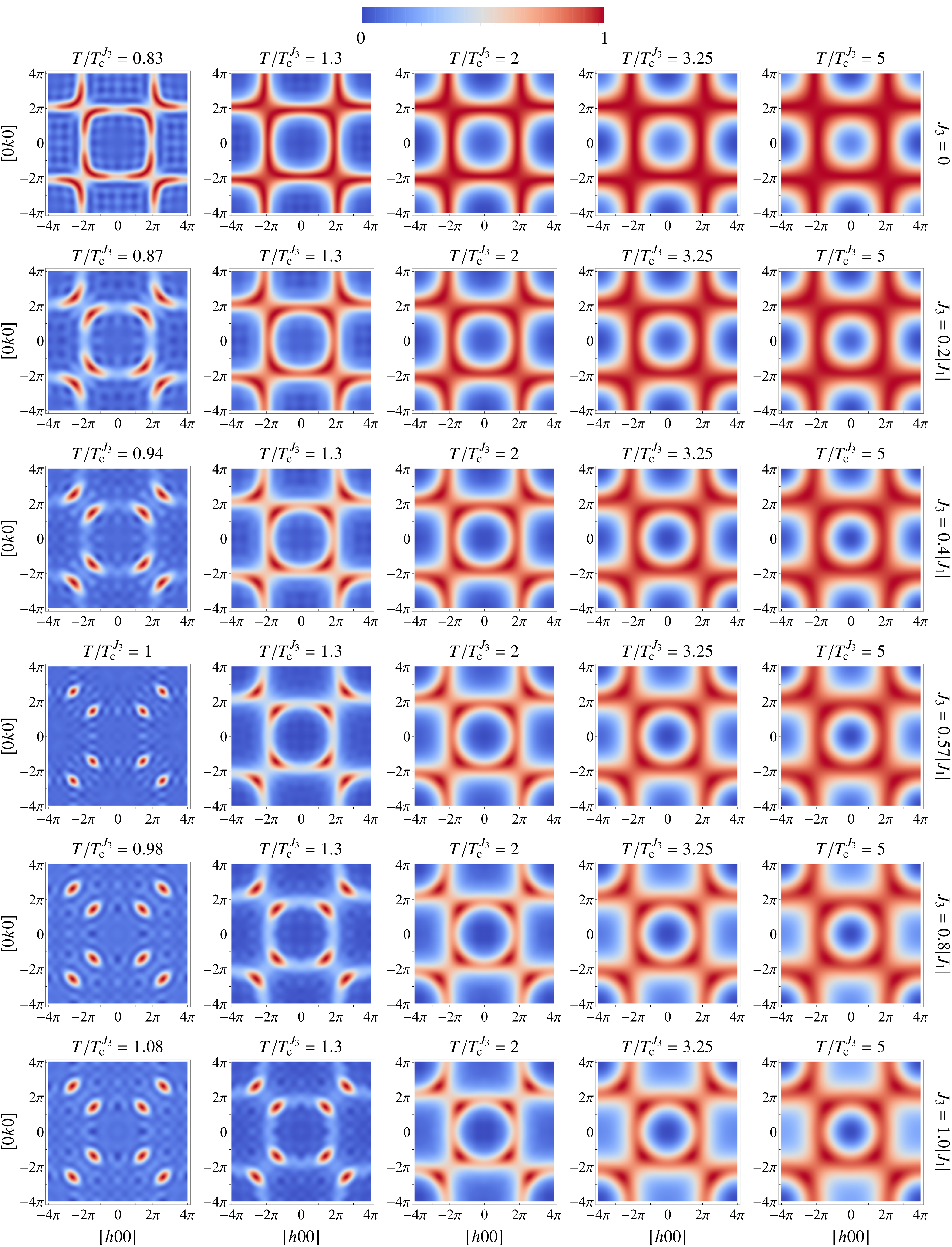}
\caption{The evolution of the spin susceptibility profile in the $q_{x}$\textendash$q_{y}$ plane with temperature for the $J_{1}$\textendash$J_{2}$\textendash$J_{3}$ Heisenberg Hamiltonian for different ratios of $J_{3}/|J_{1}|$ (different rows) keeping fixed the ratio  $J_{2}/|J_{1}|=1.64$. The ratio of $J_{3}/|J_{1}|=0.57$ corresponds to the DFT model parameters of MnSc$_2$S$_4$. The temperatures are expressed in units of the critical (ordering) temperature $T_{\rm c}^{J_{3}}$ of the model Hamiltonian with $J_{3}/|J_{1}|=0.57$. Note that at each temperature we have rescaled the susceptibility so as to make the minimum and maximum plotted values lie between 0 and 1. Corresponding to the  $J_{3}=0$ and $J_{3}=0.57|J_{1}|$ profiles, the variation of the susceptibility along the radial $(q,q,0)$ direction is shown in Fig.~\ref{fig7}.}\label{fig8}
\end{figure*}

\begin{figure*}
\includegraphics[width=0.95\linewidth]{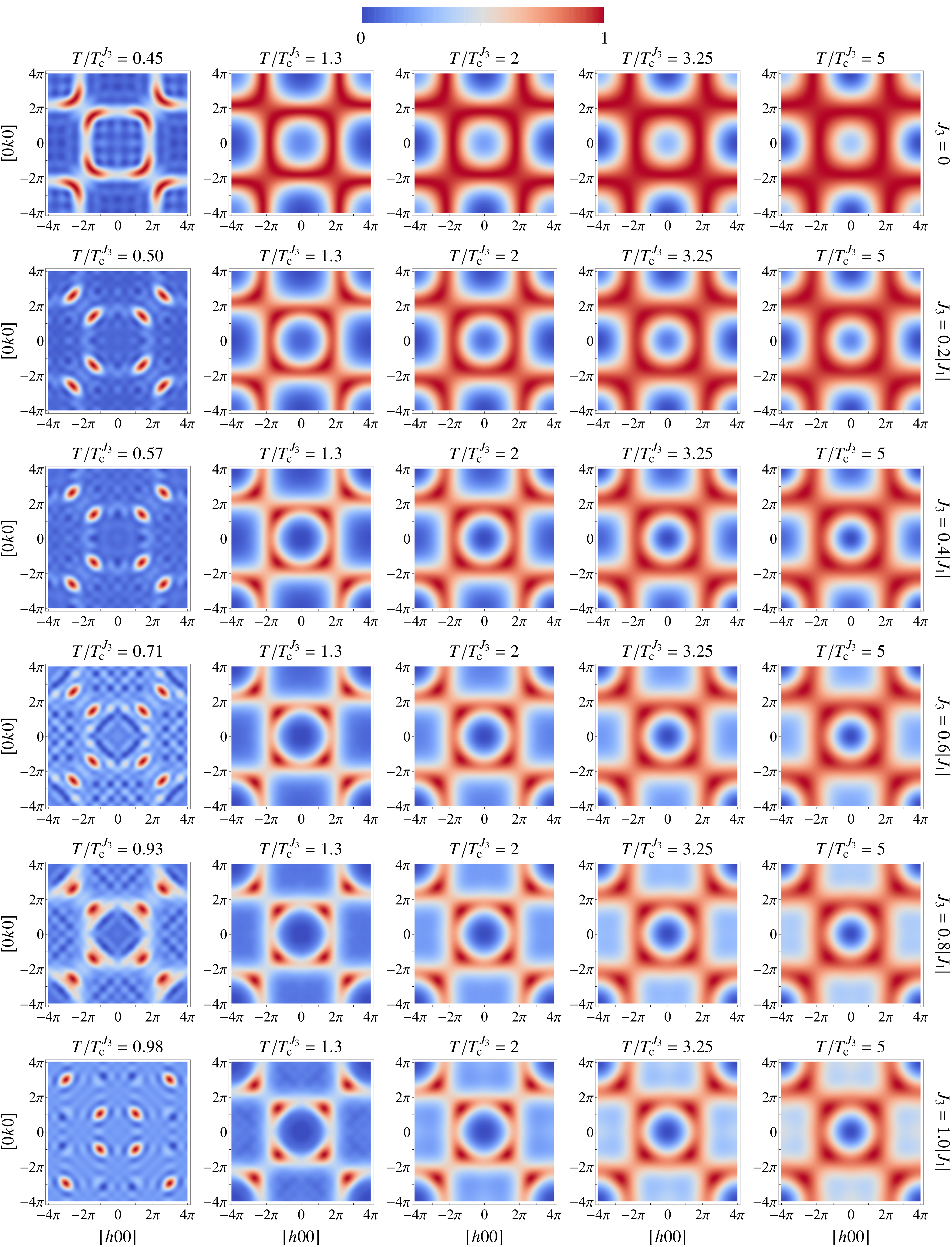}
\caption{The evolution of the spin susceptibility profile in the $q_{x}$\textendash$q_{y}$ plane with temperature for the $J_{1}$\textendash$J_{2}$\textendash$J_{3}$ Heisenberg Hamiltonian for different ratios of $J_{3}/|J_{1}|$ (different rows) keeping fixed the ratio $J_{2}/|J_{1}|=0.85$. The value of $J_{3}=0$ (first row) corresponds to an estimation of the model parameters of MnSc$_2$S$_4$ as previously determined from the radius of the spiral surface in Ref.~\cite{bergmann07}. The temperatures are expressed in units of the critical (ordering) temperature $T_{\rm c}^{J_{3}}$ of the DFT model Hamiltonian of MnSc$_2$S$_4$ with $J_{2}/|J_{1}|=1.64$ and $J_{3}/|J_{1}|=0.57$. Note that at each temperature we have rescaled the susceptibility so as to make the minimum and maximum plotted values lie between 0 and 1.}\label{fig9}
\end{figure*}

\section{Acknowledgments} 
We gratefully acknowledge the Gauss Centre for Supercomputing e.V. for funding this project by providing computing time on the GCS Supercomputer SuperMUC at Leibniz Supercomputing Centre (LRZ). The work in W\"urzburg was supported by ERC-StG-Thomale-336012, DFG-SFB 1170, and DFG-SPP 1666. J.R. is supported by the Freie Universit\"at Berlin within the Excellence Initiative of the German Research Foundation. Y.I. acknowledges the kind hospitality of the Helmholtz-Zentrum f\"ur Materialien und Energie, Berlin, where part of this work was accomplished.

\appendix

\begin{figure*}
\includegraphics[width=1.0\linewidth]{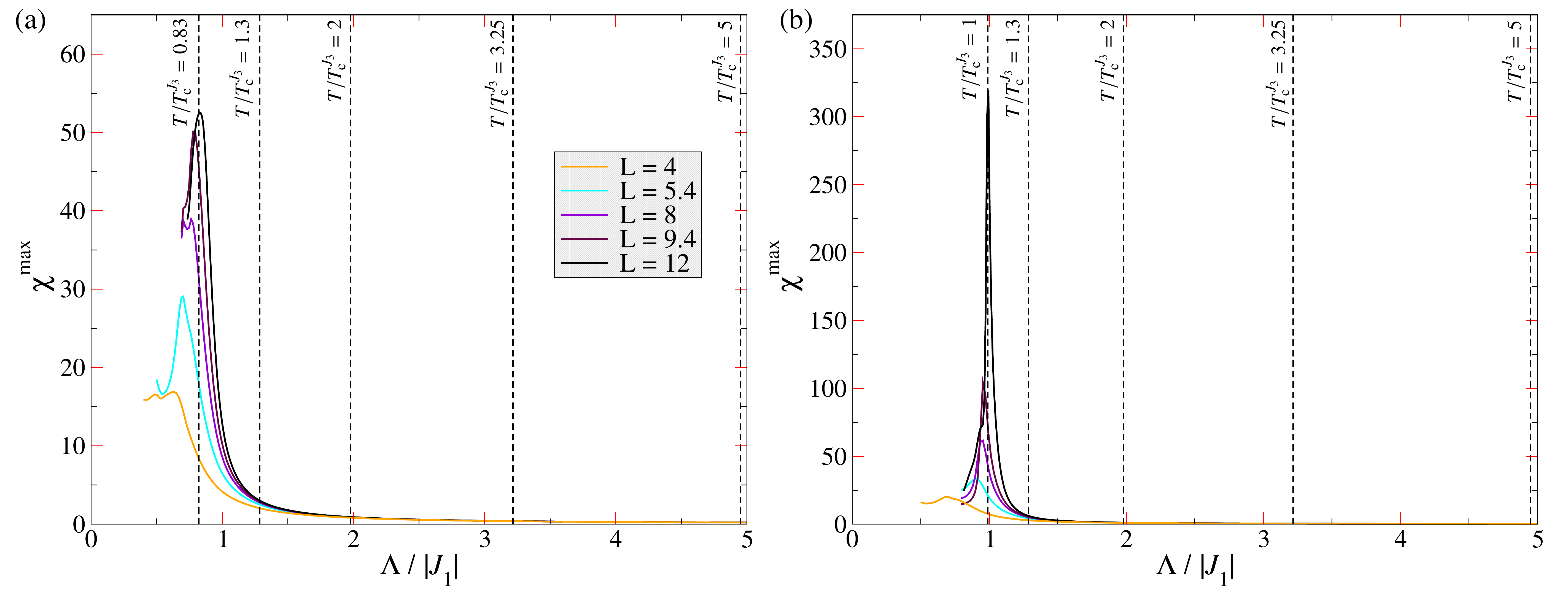}
\caption{The RG flow of the susceptibility (in units of $1/|J_{1}|$) for different ``system sizes'' wherein L is the maximal extent of the spin-spin correlator in real space in units of the nearest-neighbor distance for (a) the $J_{1}$\textendash$J_{2}$ model with $J_{2}/|J_{1}|=1.64$ and $J_{3}=0$ and (b) the $J_{1}$\textendash$J_{2}$\textendash$J_{3}$ model with $J_{2}/|J_{1}|=1.64$ and $J_{3}/|J_{1}|=0.57$.}\label{fig10}
\end{figure*}

\section{LUTTINGER-TISZA METHOD}\label{sec:LT}
The Luttinger-Tisza method aims at calculating the ground state of the classical limit of the Heisenberg model by minimizing the energy given by Eq.~(\ref{ham}), where the spin operators are substituted by classical continuous normalized vectors. To this end, the normalization of the spin vectors is replaced by the \emph{weak constraint} that the normalization only holds on average in a given spin configuration. This permits one to decompose the spin system into its Fourier modes, which is done on the two FCC sublattices of the diamond lattice separately, leading to an interaction matrix in Fourier space
\begin{equation}\label{eqn:ltmatrix}
  \tilde{J}_{\alpha \beta}(\mathbf{k}) = \sum_{i,j} J_{k} e^{\imath \mathbf{k} \cdot \mathbf{R}_{\alpha,i;\beta,j}},
\end{equation}

where $\mathbf{R}_{\alpha,i;\beta,j}$ is the vector connecting site $i$ in the FCC sublattice $\alpha$ and site $j$ in the FCC sublattice $\beta$, which are $k$th neighbors to each other. The ground state subject to the weak constraint is subsequently given by the wave vectors $\mathbf{k}$, where the lowest eigenvalue of Eq.~\eqref{eqn:ltmatrix} has its minimum. The corresponding eigenvector gives the relative weight of the mode on the sublattices, which has to have the same absolute value for a configuration to also satisfy the strong normalization constraint. Since in the diamond lattice the two sublattices are equivalent, there is no contribution proportional to $\sigma_z$ in the interaction matrix and therefore this criterion is always fulfilled, rendering the Luttinger-Tisza method exact on this lattice.

Using this method, the spiral surface shown in Figs.~\ref{fig2}(a) and \ref{fig2}(b) is obtained. Along the radial $(q,q,0)$-direction in reciprocal space the energy minimum is found at the ordering vector with
\begin{equation}
 q = 2 \arccos\left(-\frac{J_1+4 J_2+3 J_3}{4 J_2+ 8 J_3}\right)
\end{equation}
for ferromagnetic $J_1<0$.

\section{EXCHANGE COUPLINGS FROM DFT}
In Table~\ref{tab:DFT} we list the numerical values of the exchange couplings $J_1$, $J_2$, $J_3$, $J_4$ for MnSc$_2$S$_4$ obtained from DFT [see also Fig.~\ref{fig1}(b)]. The couplings have been calculated for five different values of the Hubbard interaction ranging from $U=3$ eV to $U=5$ eV. Also shown is the Curie-Weiss temperature $\Theta_\text{CW}$ for each set of spin interactions. In our PFFRG calculations we use the parameters corresponding to $U=4.5$ eV since this leads to the best agreement of the Curie-Weiss temperature with the experimental value $\Theta_\text{CW}=-23$ K. 

\section{TEMPERATURE EVOLUTION OF THE SUSCEPTIBILITY}
In Fig.~\ref{fig7} we show the susceptibility along the radial $(q,q,0)$ direction for different temperatures where the coupling parameters are the same as in Fig.~\ref{fig4}. For each plotted temperature the susceptibility is normalized with respect to its value at $(2\pi,2\pi,0)$ to compensate for an overall decrease with temperature. Our results for $J_3=0$ [Fig.~\ref{fig7}(a)] and $J_3=0.57|J_{1}|$ [Fig.~\ref{fig7}(b)] both show a clear broadening of the susceptibility along the radial $(q,q,0)$ direction as temperature increases; see also the inset of Fig.~\ref{fig5} (the oscillating behavior of the red curves at small susceptibilities is an artifact caused by the finite number of Fourier components included in our numerics). At small temperatures the susceptibility shows a clear double peak structure, where the peak at smaller $q$; i.e., $q/2\pi<1$ belongs to the spiral surface centered around $(0,0,0)$ and the peak with larger $q$, i.e., $q/2\pi>1$, corresponds to the spiral surface centered around $(4\pi,4\pi,0)$. A pronounced double peak indicates that different spiral surfaces are clearly distinguishable, pointing towards an intact spiral spin liquid. As can be seen in Fig.~\ref{fig7}, with increasing temperature, the two peaks smear out considerably faster for $J_3=0$ as compared to $J_3=0.57|J_{1}|$, implying that a finite $J_3$ coupling may also aid in stabilizing a spiral spin liquid. The most pronounced peak structure is observed for $J_3=0.57|J_{1}|$ close to criticality [red curve in Fig.~\ref{fig7}(b)]. In this case, however, a strong selection of spiral states along the surface takes place [see Figs.~\ref{fig3}(f) and \ref{fig5}], indicating the onset of conventional long-range magnetic order instead of the formation of a spiral spin liquid. We have also investigated the temperature evolution of the susceptibility profile for different values of $J_{3}$ so as to systematically study the role of a $J_{3}$ coupling. The results for a model with fixed $J_{2}/|J_{1}|=1.64$ and varying $J_{3}/|J_{1}|=0, 0.2, 0.4, 0.57, 0.8, 1$ are shown in Fig.~\ref{fig8}, while results for a model with fixed $J_{2}/|J_{1}|=0.85$ and varying $J_{3}/|J_{1}|=0, 0.2, 0.4, 0.6, 0.8, 1$ are shown in Fig.~\ref{fig9}. A few trends are worth noticing. (i) The critical (ordering) RG scale is found to increase with increasing $J_{3}$, pointing to the fact that third neighbor interactions relieve the frustration. (ii) The spiral selection effect becomes progressively more pronounced with increasing $J_{3}$, and consequently the intensity smearing with increasing temperature occurs at a slower pace, such that the spiral surface is recovered at progressively higher temperatures with increasing $J_{3}$. (iii) The selection remains always of the $(q,q,0)$ wave vector type. 

\section{FINITE-SIZE EFFECTS}
In Fig.~\ref{fig10} we show PFFRG results for (a) $J_{3}=0.57|J_{1}|$ and (b) $J_{3}=0$ when varying the system size (i.e., when varying extent of the spin correlations in real space). We observe that the critical ordering scale increases upon increasing the system sizes; however, it appears to converge to a good degree of accuracy for the largest system sizes we have simulated. Nonetheless, at higher temperatures, i.e., $T/T_c \geqslant 1.3$, which is the value used for comparison with experiments and is relevant for observing the spiral spin liquid, the PFFRG results have already converged.

\end{document}